\documentstyle[aps,prl,preprint,floats,epsfig]{revtex}
\begin{document}

\tighten

\title{
{\tighten \vskip -1cm \small 
\begin{flushright}
CLNS 98/1551   \\
CLEO 98-6
\end{flushright} }
A limit on the mass of the $\nu_\tau$ }

\author{CLEO Collaboration \\
(\today) }

\maketitle

\begin{abstract} 
A limit
 on the mass of the tau neutrino $m_{\nu_\tau}$ is derived from  
$4.5 \times 10^6$ tau pairs produced in an integrated luminosity of 
  5.0 $fb^{-1}$ of  
$e^+e^- \to \gamma* \to
 \tau^+ \tau^-$ reactions at center of mass energies $\approx$ 10.6 GeV. 
 The measurement technique involves a two-dimensional extended likelihood
  analysis, including  the dependence of the end-point population
 on $m_{\nu_\tau}$, and allows for the first time an explicit
 background contribution. We use the 
 the decays $ \tau \to 5 \pi \nu_\tau$ and $\tau \to 3 \pi 2\pi^0 \nu_\tau$ 
 to obtain  an upper  limit of 30 MeV/$c^2$ at $95\%$ C.L.
\end{abstract}

\newpage

\begin{center}
R.~Ammar,$^{1}$ P.~Baringer,$^{1}$ A.~Bean,$^{1}$
D.~Besson,$^{1}$ D.~Coppage,$^{1}$ C.~Darling,$^{1}$
R.~Davis,$^{1}$ S.~Kotov,$^{1}$ I.~Kravchenko,$^{1}$
N.~Kwak,$^{1}$
% ...
\end{center}
 
\small
\begin{center}
$^{1}${University of Kansas, Lawrence, Kansas 66045}\\
% ...
\end{center}

\newpage

 Although the tau neutrino has never been directly observed, the question
 of its mass is an important issue in  particle physics and cosmology.
 While the requirement that the density of primordial relic neutrinos
 from the
 Big Bang  not
 over-close the Universe restricts the mass of a stable neutrino to
 be less than $\sim$ 100 eV/$c^2$~\cite{Cowsik,Kolb,Peebles},
  unstable neutrinos are less 
 restricted.
 Big Bang nucleosynthesis models allow for a massive neutrino in the
 10 to 31 MeV/$c^2$ range for lifetimes in the 0.01 to 40   second 
 interval~\cite{kawasaki}\footnote{
 Recently~\cite{rehm} it has been pointed out that these arguments do depend upon
 which set of astrophysical observations are used.}.
 This astrophysically allowed range currently overlaps the experimentally accessible
 bounds.
 The most stringent accelerator-based limits on $m_{\nu_\tau}$ are 
 derived from one-dimensional mass fits and 
 two-dimensional  energy versus mass fits  of the hadrons in tau decays. 
 The sensitivity of the method is largest
 when the invariant hadronic mass is large, and there is little phase
 space for the (unseen) neutrino such as in the decays modes $\tau \to 5\pi(\pi^0) \nu_\tau$ and
 $\tau \to 3\pi 2 \pi^0 \nu_\tau$. In addition, due to its large branching ratio,  the
 decay $\tau \to 3\pi  \nu_\tau$ can be  used in spite of its lower inherent sensitivity per event
 resulting from its low average hadronic decay product mass.

 The lowest published upper limit\footnote{All neutrino mass upper limits quoted herein
 are at the 95$\%$ confidence level.}, by the
 ALEPH collaboration~\cite{ALEPH}, is 24 MeV/$c^2$ 
 and is obtained
 with a two-dimensional likelihood fit to a region near the endpoint
 of 25  $  5 \pi (\pi^0) $ candidate events. 
 By doubling this dataset and adding in $ 3\pi $ decay 
 candidates,
 the ALEPH collaboration~\cite{ALEPH2}
 has recently reported an upper limit of 18.2 MeV/$c^2$.
 The DELPHI collaboration~\cite{DELPHI}  has claimed a limit of 
 33 MeV/$c^2$ in the  $3\pi$ mode from a  mass fit, but notes that
 this limit is sensitive to the contribution of a possible higher
 mass resonance in the $3\pi$ mass distribution. DELPHI estimates
 that the model dependence in the $3\pi$ mode decreases the limit
 to 62 MeV/$c^2$. 
The OPAL collaboration~\cite{OPAL2}
  quotes a
 35.3 MeV/$c^2$ limit from a two-dimensional fit to the missing
 momentum and missing mass in $ 3 \pi $ 
 decays using the event thrust axis as an estimator of the tau direction.
 Combining this with an earlier limit of 74 MeV/$c^2$ from 
 $  5 \pi $ decays~\cite{OPAL1}, a 29.9 MeV/$c^2$ limit 
 is obtained.  ARGUS~\cite{ARGUS1,ARGUS2}
  has published a 31 MeV/$c^2$ limit based on a one-dimensional mass fit to
 20 $  5 \pi $ events. The previous CLEO limit~\cite{CLEO1} of 32.6 MeV/$c^2$
 was set with a fit to the mass spectrum of 60 $  5 \pi $ events, and
 53 $ 3 \pi 2\pi^0 $ events. 

The current analysis uses data collected with
 the CLEO II detector~\cite{cleodet} at the Cornell $e^+e^-$ storage
 ring (CESR) at center of mass energies near 10.6 GeV. The data set 
 consists of  5.0 $fb^{-1}$ of integrated luminosity corresponding to  4.5 million produced
 tau pairs. The
 tau's, created mainly via $e^+e^- \to \gamma* \to
 \tau^+ \tau^-$, have the full beam energy of 5.29 GeV 
 and recoil back to back, modulo initial state radiation
 effects. 
 The decay modes studied
 are $ \tau \to 5 \pi \nu_\tau$ and $\tau \to 3 \pi 2\pi^0 \nu_\tau$ 
recoiling against a leptonically-decaying tau (the ``tag'').
 An upper limit on $m_\nu$ is obtained from a two-dimensional likelihood
fit to hadronic mass and energy in these decays. The likelihood
 includes for the first time a background contribution, as well as
a factor describing the expected number of events as a function of
 neutrino mass, and measured event errors for hadronic mass and energy.

In order to avoid a spuriously low limit from a background event, 
 the event selection criteria
 are chosen to obtain a very  pure  event sample of $ \tau \to 5 \pi \nu_\tau$ or 
 $3\pi2\pi^0\nu_\tau$, and roughly
 follow those of our last publication~\cite{CLEO1}. 
A clean sample of tau's is obtained by selecting events in
 which the tag decays leptonically:  $\tau \to 
 e \nu\nu_\tau$ or $\tau \to  \mu\nu\nu_\tau$. We thus require the 
 event to contain an isolated  
 lepton ($e$ or $\mu$) in the  barrel ($|\cos{\theta}| < 0.71$) recoiling against
 5 or 3 charged tracks, with event hemispheres defined by the event
 thrust axis.
 Exactly 6 or 4 charged tracks consistent with production at the nominal
 interaction point are required, after  pairs of tracks
 consistent with photon conversions and Dalitz $\pi^0$ or $K_S^0$ decays
 are rejected. 
 Backgrounds from two photon events are minimized
 by requiring that the total  missing momentum point away from the
 beamline, and that the total energy transverse to the beam axis 
 be above 200 MeV.  
 To supress hadronic events, the total 
 visible energy is required to be below 90\% of the total center of mass 
 energy.
 All hadronic tracks  must also have a 
 specific ionization ($dE/dx$) consistent
 with that of a pion. We correct for $dE/dx$ energy loss
when computing energies and invariant masses.

 Isolated showers in the calorimeter with photon-like lateral 
 shower profiles
 are identified as photons.
 In the barrel, photon candidates must have an 
 energy above
 60 MeV, while in the range $0.71  < |\cos{\theta}| < 0.95 $, they
 must have an energy above 100 MeV. Exactly 0 or 2 neutral pions must
 be reconstructed on the signal side, with none on the
 tag side.  These $\pi^0$ candidates must have an  
 opening angle of less than $60^\circ$ and 
 a  mass constrained fit of the $\pi^0$ candidates must have a confidence 
 level greater than 5\%. If the photon candidates can
 be combined into $\pi^0$s in multiple ways, we select the combination
 with the smallest total $\chi^2$ for the $\pi^0$ mass hypothesis.
 We permit calorimeter showers that are unassociated with charged 
 tracks or $\pi^0$s to remain. 
 The summed  energy of all showers within $90^\circ$ of the lepton
 direction must be less than 100 MeV in each of the barrel and endcap region of the calorimeter.
 For the signal hemisphere, the energy requirement is increased to 200 MeV.

 The selection criteria above were designed to minimize backgrounds, which
 arise primarily  from non-tau decays such as $e^+e^-\to q\overline{q}$ 
 or  $e^+e^-\to \Upsilon(4S) \to B\overline{B}$ or from 
 feedthrough from other tau decay modes. A data sample which satisfies the 
 above selection  criteria on the signal side of the event, but with an
  invariant 
 mass above  the tau mass on the tag side, is used as an estimator of 
 non-tau backgrounds. 
 The background estimate is obtained by scaling the number of these estimator 
 events above the tau mass on the signal side to the number of data 
 events above the tau 
 mass on the signal side minus the Monte Carlo predicted number of taus
 misreconstructed above the tau mass.
 The background estimates obtained in this way are consistent with those
 obtained by using $e^+e^- \to q\overline{q}$ Lund~\cite{lund} generated 
 Monte Carlo events.
 Monte Carlo events with full detector simulation~\cite{geant} based on the 
 event generators KORALB~\cite{koralb} and PHOTOS~\cite{photos} were used to estimate 
 tau decay 
 backgrounds in the signal hemisphere (feed-across).

\begin{figure}[ht]
\centerline{
\epsfig{file=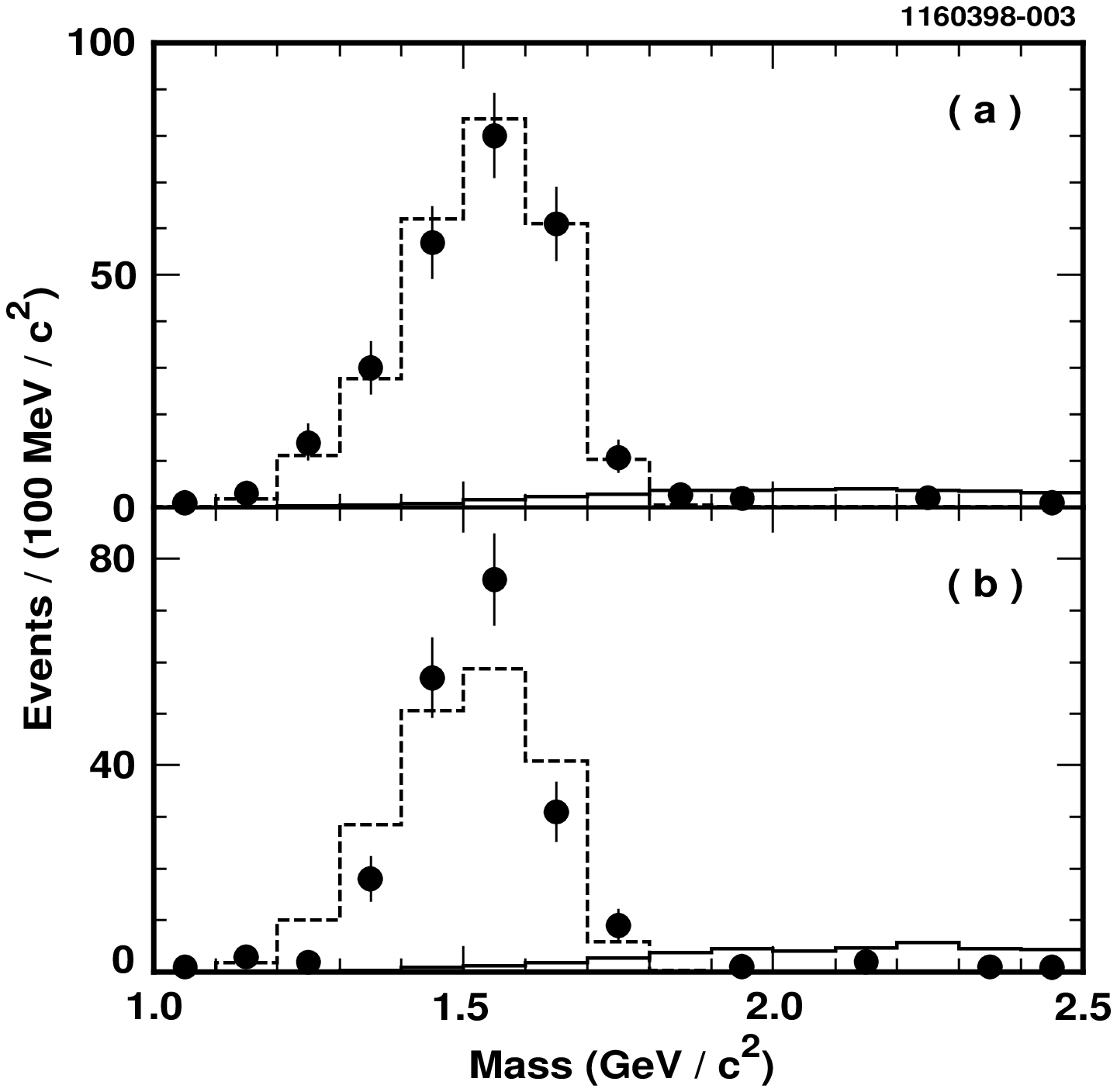,width=0.95\textwidth }}
\caption{ The mass distributions of the $5\pi$ (a) and 
 $3\pi2\pi^0$ (b) data samples. 
 The dashed histogram represents the shape expected from Monte Carlo
 normalized to the number of events below the tau mass.
 The background
 estimate is shown as a solid histogram and is normalized for display
 purposes to five times its nominal value. }
\label{fig:mass}
\end{figure}

 The resulting event distributions, shown in Fig.~\ref{fig:mass}, contain
 266  events in the $5\pi$ final state, of which 8 are
 above the tau mass. The $3\pi2\pi^0$ mode contains 207 events, of which
 13 are above the tau mass.
 Monte Carlo simulations show a reconstruction
 efficiency of $(3.08 \pm 0.10 )\%$  for the $5\pi$ mode
 and $(0.43 \pm 0.02 )\%$ for the
 $3\pi2\pi^0$ mode.  
 In both modes the number of events observed
 is consistent with the expectation from the 
  world average branching fractions~\cite{pdg96}.

\begin{figure}[ht]
\centerline{
\epsfig{file=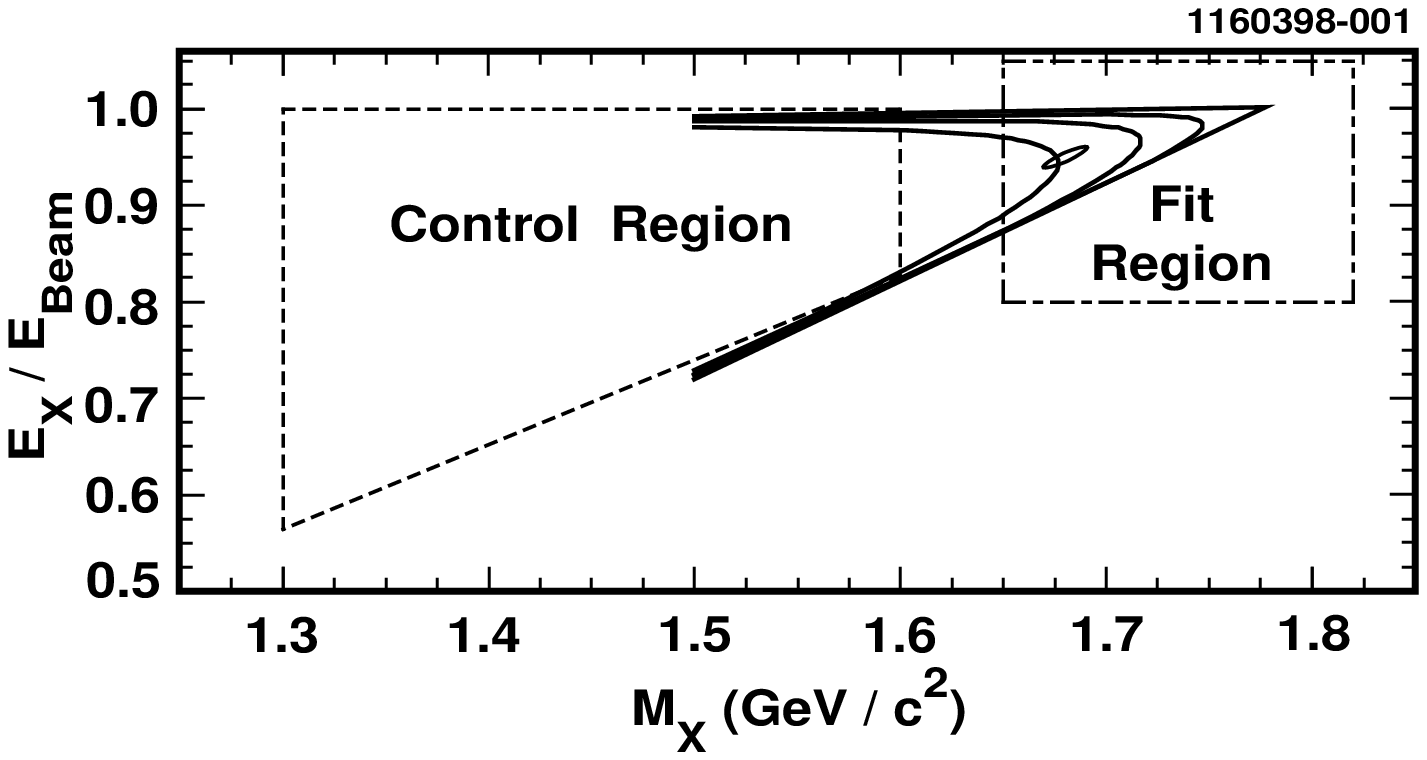,width=0.95\textwidth }}
\caption{ The number of events in the 
 the fit region $\Omega$  relative to the number in the
  control region ${\cal R }$ 
  in the scaled hadronic energy versus hadronic
 mass plane  is  a function of neutrino mass.
 Kinematically allowed neutrino mass contours are 
 drawn for neutrino masses of 0, 30, 60 and 100 MeV/$c^2$. 
 Note the typical error
 ellipse drawn in the fit region. }
\label{fig:fitregion}
\end{figure}

\begin{figure}[ht]
\centerline{
\epsfig{file=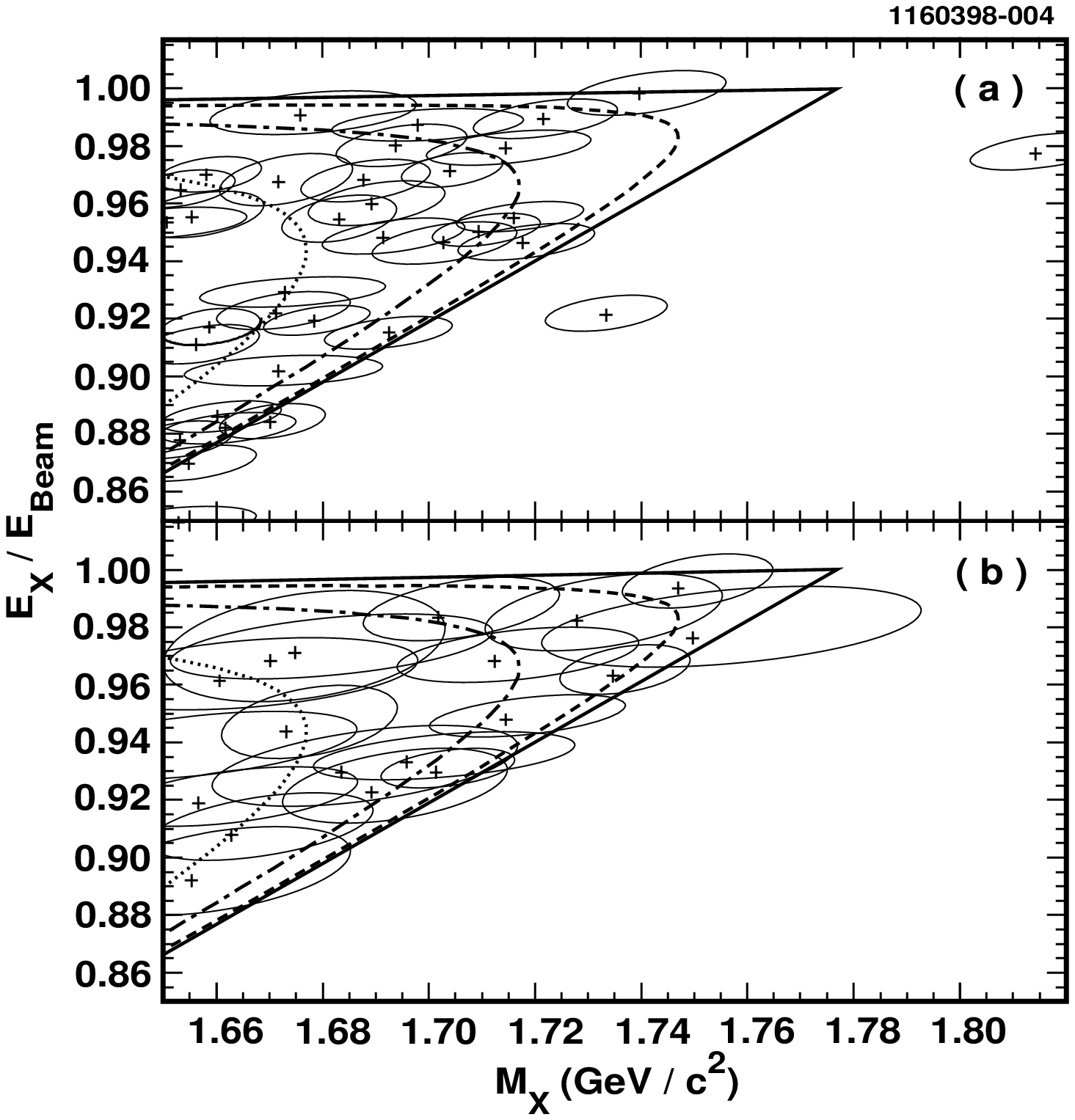,width=0.95\textwidth }}
\caption{The hadronic scaled energy versus mass distribution of the 
$5\pi$ (a)  and $3\pi2\pi^0$ (b) data samples
in the fit region. Ellipses represent 1 $\sigma$ resolution errors 
and incorporate reconstruction induced systematic offsets. Kinematically
 allowed contours are  drawn for neutrino masses of 
0, 30, 60 and 100 MeV/$c^2$ as solid, dashed, dot-dashed and 
dotted lines respectively. Events below the kinematically allowed
 region are fully consistent with signal events, once initial
 state radiation effects are considered (as they are in the likelihood.)
 }
\label{fig:evsm}
\end{figure}

Each signal event is represented by a point in the two-dimensional plane
 formed by the hadronic energy scaled to the beam energy ($E_X/E_B$) versus 
 hadronic mass ($M_X$), 
with the estimated errors represented by an ellipse. 
 The sensitivity to neutrino
 mass is largest in the region near $M_X=m_\tau$ and  $E_X/E_B=1$, as shown
 in Fig.~\ref{fig:fitregion}.
 The data events are shown in Fig.~\ref{fig:evsm}.
We determine the tau neutrino mass using the events lying 
  in the fit region $\Omega$ shown in Fig.~\ref{fig:fitregion}. 
 There are 36 $5\pi$ events in the fit region, with negligible tau backgrounds
 and an estimated
 0.3  events from non-tau backgrounds. There are
 19 $3\pi2\pi^0$ events in the fit region, with 1.0  of these expected as
 feed-across from other tau decay modes and 0.4 expected from non-tau 
 backgrounds. 
 Details of the analysis are summarized in Table~\ref{tab:summary}.
\begin{table}
\caption{ Summary of signal, background, efficiency, resolution and
 upper limit by mode. }
\begin{center}
\begin{minipage}[t]{12cm}
\label{tab:summary}
\begin{tabular}{|| c | c | c ||} \hline
   	Mode  			&  \hskip -5cm $5\pi$ 
 &  \hskip -5cm $3\pi 2\pi^0$   \\  \hline\hline
   Total Events  		&  \hskip -5cm 266 	
& \hskip -5cm 207 \\ \hline
   Events in Fit Region 	&  \hskip -5cm 36
& \hskip -5cm 19 \\ \hline
   Signal Region Purity ($\%$) 	& \hskip -5cm 99 
& \hskip -5cm 93 \\ \hline
   Selection Efficiency ($\%$)	&  \hskip -5cm 3.08
& \hskip -5cm 0.43  \\ \hline
Typical Mass Resolution (MeV/$c^2$) &  \hskip -5cm 15
&  \hskip -5cm 25 \\ \hline
Typical Energy Resolution (MeV) &  \hskip -5cm 25
&  \hskip -5cm 50 \\ \hline
Upper Limit $@$ 95 $\%$ CL  (MeV/$c^2$) &  \hskip -5cm 31
&  \hskip -5cm 33 \\ \hline
\end{tabular}
\end{minipage}
\end{center}
\end{table}

 We extract a measurement of the tau neutrino mass using an 
 unbinned extended maximum likelihood technique.
 The procedure fits for one parameter ($m_\nu$), taking as input the 
 measured hadronic energy and  mass
 of the events in the fit region shown in Fig.~\ref{fig:fitregion}. 
 The likelihood function, detailed in the appendix, is composed of 
 a Poisson factor, expressing the number of events expected, 
 times the sum  
 of a signal term  and a background term. The likelihood is calculated
 as a function of neutrino mass using a novel technique. Instead
 of using explicit parameterizations of the Monte Carlo in a
 likelihood convolution~\cite{LUCA}, 
  Monte Carlo signal events are used in the evaluation
 of the likelihood, directly implementing the best knowledge of
 all physics effects including initial state radiation and
 detector acceptance. The only explicitly parameterized term is 
 the detector smearing function.

\begin{figure}[ht]
\centerline{
\epsfig{file=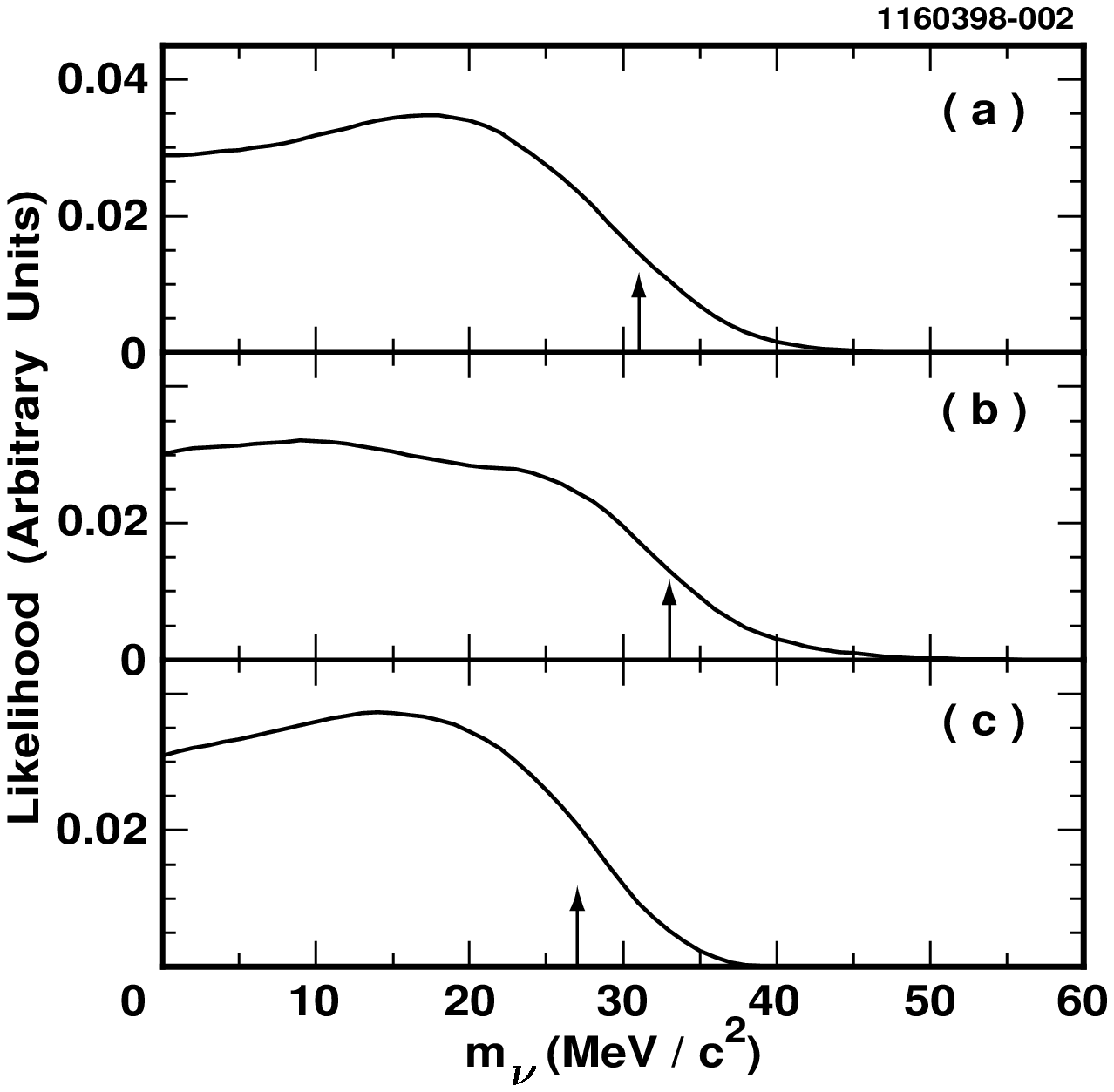,width=0.95\textwidth }}
\caption{The likelihood as a function of neutrino mass
 for the $5\pi$ (a), $3\pi2\pi^0$  (b) and combined (c)
 data samples.
 The 95\% CL upper limits, uncorrected for systematic errors, 
  are at 31, 33, and 27 MeV/$c^2$, 
 respectively.
 }
\label{fig:likeli}
\end{figure}

 The resulting extended likelihood is shown
 in Fig.~\ref{fig:likeli}.
We define\footnote{The method for extracting an upper limit from a likelihood distribution
at a given confidence level is not unambiguously defined; the method
used here differs from that used in the analysis of Ref.~\cite{ALEPH}. Therefore
comparisons of upper limits among different experiments must be done
with care.}
 the  95\% confidence level (CL) upper limit by integrating
 defined likelihood above zero mass to its ninety-fifth percentile.
 We find
 95\% CL upper limits of 33, 31, and 27 MeV/$c^2$ for the $3\pi2\pi^0$,
 $5\pi$ and combined samples. A separate analysis using only mass 
 information yields a combined upper limit of 31 MeV/$c^2$.

 The interpretation of this limit as a meaningful statement about 
 probability is conditional upon the measured likelihood being
 representative of an ensemble of similar experiments. 
 The event distribution and the observed number of events at the
 endpoint are consistent with our Monte Carlo estimation with
 zero neutrino mass, using the spectral function tuned to the
 data as described below.
An ensemble of Monte Carlo experiments using statistics 
compatible with those we observe reveal that
 a smaller upper limit is obtained in 67\% of  experiments with a massless
 neutrino. The average expected upper limit given our sample size is
 found to be 25 MeV/$c^2$.

 Following the conservative  prescription used by
 the LEP experiments~\cite{LUCA}, a linear systematic error, $\Delta(m_{95})$,
  is added to the limit. The error is defined by 
 $$\Delta(m_{95})^2 = \Sigma (\tilde{m}_{95} - m_{95})^2 ,$$ 
 where the sum is
 over systematic error sources, and the difference represents the change
 in the nominal upper limit from each error source. 
 Systematic errors are summarized in Table~\ref{tab:syst}.
\begin{table}
\begin{center}
\begin{minipage}[t]{7cm}
\caption{ Systematic error summary. }
\label{tab:syst}
\begin{tabular}{|| c | c ||} \hline
    Source     &  \makebox[2.9cm][l]{Error  (MeV/$c^2$)}  \\  \hline\hline
   Spectral Function & \makebox[1cm][l]{1.9}  \\ \hline
   Smearing Tails    & \makebox[1cm][l]{1.6} \\ \hline
   Mass Scale & \makebox[1cm][l]{1.1} \\ \hline
   Smearing Parameters & \makebox[1cm][l]{0.7}  \\ \hline
   MC Statistics & \makebox[1cm][l]{0.6} \\ \hline
   Background & \makebox[1cm][l]{0.4} \\ \hline
   Energy Scale & \makebox[1cm][l]{0.3} \\ \hline \hline
   Total & \makebox[1cm][l]{2.9} \\    \hline \hline
\end{tabular}
\end{minipage}
\end{center}
\end{table}

The largest systematic error 
 is due to  uncertainties in the spectral functions. We obtain 
 the spectral functions from a fit
 to a statistically independent sample of decays recoiling against
 pionic decays of the tau, excluding the 100 MeV/$c^2$ near the tau mass 
 and including
 a background term. Each fit is  a sum of two 
 phenomenologically inspired spectra~\cite{Tsai,Gilman,Pham};
 the dominant spectrum  is the distribution
 predicted for the  5 pion decay of the tau as derived from the 
 CVC predictions for 4 pion decay and soft pion theorems, and the other 
 is the harder spectrum CVC predicts for 6 pion decays. 
 The systematic variation
 for the spectral function is taken as the allowed variation in the fit
 parameters and results in errors of 1.5 MeV/$c^2$ from the $5\pi$  mode and 1.1 MeV/$c^2$ from 
 the $3\pi2\pi^0$ mode.

 The next largest systematic error comes from  the modeling of the
 detector smearing function.
 The  smearing function  is approximated by
 a sum of three two-dimensional Gaussians to model its extended tails. 
 The smearing function parameters are derived from the Monte Carlo 
 simulation, while the
 widthes of each Gaussian are proportional to the 
 event error, as estimated from the propagated tracking error matrix 
 and  the photon reconstruction errors.
 The errors from the modeling of the tails of the smearing function
 are estimated by using a fourth Gaussian component in separate
 fits to  charm and $B$ decay  data. 
 The mass (energy) smearing is 
 studied with charm ($B$) decays to final states with 
 similar numbers of tracks as our samples.
 $B$ decays are used since the $B$, created nearly at rest in the CLEO detector
 frame, 
 allows the comparison of reconstructed energy with the beam energy. 
 This
 smearing uncertainty is 1.4 MeV/$c^2$ for the $ 3\pi 2\pi^0 $ mode and 0.6 MeV/$c^2$ 
 for the $5\pi$ mode. 
In addition, all smearing function parameters are allowed to
 vary within their fitted uncertainties, resulting in another 
 0.5 MeV/$c^2$ uncertainty for each of the two modes.

 The uncertainty in the absolute mass and energy scales of 1 MeV/$c^2$
 (as estimated from charm decays)
 and 2.7 MeV (from $B$ decays), respectively, contribute
 1.1 MeV/$c^2$ and 0.3 MeV/$c^2$ to the final systematic error estimate.

 Other sources of error considered are the Monte Carlo statistics used in
 the likelihood evaluation and the  background size. 
 The total systematic error from all sources added in quadrature is 2.9  MeV/$c^2$, 
 resulting in a final 95\% C.L. upper limit on the mass of the 
 tau neutrino of 30 MeV/$c^2$.

 The use of the Poisson term in this analysis' likelihood
 is an improvement over a previous technique~\cite{ALEPH} because it
 reduces the variance in fitted neutrino mass, thus
 decreasing the probability of a spuriously low limit in the presence
 of a massive neutrino. 
 The present analysis also uses measured event 
 errors to avoid a possible bias.
 The smearing shape used in previous analyses~\cite{ALEPH,OPAL1}  
 depends on repeated Monte Carlo simulation of the data points, 
 using the  measured values as input to the Monte Carlo event generator. 
 Such a method neglects the fact that events reconstructed near the
 endpoint stand a larger chance of being upward fluctuations
 from low mass events than do events in the middle of the accepted 
 distribution, and thus tends to underestimate the mass and energy 
 errors associated with each data point.

 In summary, using the world's largest sample of candidate
 $ \tau \to 5 \pi \nu_\tau$ and $\tau \to 3 \pi 2\pi^0 \nu_\tau$ decays, 
we obtain a 95\% confidence 
level upper limit on $m_{\nu_\tau}$ of 30 MeV/$c^2$. The method used properly takes
 into account the change in the number of events expected at the endpoint
 as a function of $m_{\nu_\tau}$,
 as well as  a background contribution, and a detailed detector smearing 
 function.

We gratefully acknowledge the effort of the CESR staff in providing us with
excellent luminosity and running conditions.
J.R. Patterson and I.P.J. Shipsey thank the NYI program of the NSF, 
M. Selen thanks the PFF program of the NSF, 
M. Selen and H. Yamamoto thank the OJI program of DOE, 
J.R. Patterson, K. Honscheid, M. Selen and V. Sharma 
thank the A.P. Sloan Foundation, 
M. Selen thanks Research Corporation, 
S. von Dombrowski thanks the Swiss National Science Foundation, 
and H. Schwarthoff thanks the Alexander von Humboldt Stiftung for support.  
This work was supported by the National Science Foundation, the
U.S. Department of Energy, and the Natural Sciences and Engineering Research 
Council of Canada.

\appendix
\section*{Appendix}
 The likelihood is given by:
\begin{eqnarray*}
 {\cal L}( m_\nu ) = {\cal P }\left( N_{obs} \right)
        \prod_{Events}
         {\cal L}_{dat}( \tilde{X}_{data}| \sigma_{data}, m_\nu,\overline{N}_{bgd}).
 \end{eqnarray*}

\noindent The first factor is the Poisson probability of the number of events 
 observed in the fit region, $N_{obs}$. 
 The second factor is the product likelihood over the data 
 events observed $\tilde{X}_{data}$ , and their associated errors 
 $\sigma_{data}$, given an expected number of background events,
$\overline{N}_{bgd}$.

 The Poisson factor is 
\begin{eqnarray*}
{\cal P }\left( N_{obs} \right) =
{\left( \overline{N}\left(m_\nu\right)+\overline{N}_{bgd}\right) }^{N_{obs}}
\left[
e^{-\left(\overline{N}\left(m_\nu\right)+\overline{N}_{bgd}\right) }
 \over{N_{obs}!}
\right].
\end{eqnarray*}
 \noindent The expected number of signal events in the fit region $\Omega$,
 $\overline{N}\left(m_\nu\right)$,
is a 
 function of neutrino mass and
 is extrapolated from the number of events seen in the control region
 $\cal R $  using the partial width:
\begin{eqnarray*}
\overline{N} \left(m_\nu\right)  =
 { \int_\Omega d\Gamma\left(m_\nu\right) \over{
       \int_{\cal R } d\Gamma\left(m_\nu = 0\right)}} 
          N^{\cal R }_{data}.
\end{eqnarray*}

\noindent The control region ${\cal R}$  is chosen away from the neutrino
 mass sensitive region so that the number of events expected
 in it       
 does not depend strongly on the neutrino mass 
(see Fig.~\ref{fig:fitregion}). The partial width is fit from an independent
 data sample as 
  described in the text.

 The event likelihood can be expanded as:
\begin{eqnarray*}
 {\cal L}_{dat}( \tilde{X}_{data}| \sigma_{data}, m_\nu,\overline{N}_{bgd} )
  =  
      (1-{\overline{N}_{bgd}\over{N^\Omega_{obs}}} )  
              {\cal L}_{sig}( \tilde{X}_{data}| \sigma_{data}, m_\nu ) \\
    + ({\overline{N}_{bgd}\over{N^\Omega_{obs}}} )  
              {\cal L}_{bgd}\left(\tilde{X}_{data}| \sigma_{data}\right) .
\end{eqnarray*}
\noindent The first term of this expression is the neutrino mass likelihood
 for a pure signal. The second term expresses the background shape.
 The likelihoods ${\cal L}_{sig}$ and ${\cal L}_{bgd}$
 are normalized over the fit
 region $\Omega$ and neutrino mass. The notation $\tilde{X}_{data}$ 
 is to stress that
 we use the reconstructed (smeared) variables for the data.

 The background likelihood, ${\cal L}_{bgd}$, is a parameterization of the background event
 distribution from the high tag-mass sample described in the text.

 The signal likelihood, ${\cal L}_{sig}$, is determined from Monte Carlo simulations, taking
  into account detector smearing,
 the physics of tau decays, and initial state radiation
 effects.  The analytic expression for this 
 likelihood, as used in previous analyses (see for instance~\cite{LUCA}),
  can be written as:
\begin{eqnarray*}
{\cal L}_{sig} ( \tilde{X}_{data}|\sigma_{data} , m_\nu ) & = & 
{ \int_{All \,X }{dX \,  G(X,\tilde{X}_{Data},\tilde{\sigma}_{Data})
\, \epsilon(X)\, f_{Phys}(X|m_\nu) }\over{ {{\cal N}(m_\nu)}}},
 \\ 
{\cal N}(m_\nu)& = & \int_{\Omega} d\tilde{X}\int_{All \,X }
{dX \,  G(X,\tilde{X},\tilde{\sigma})\, \epsilon(X)}\, f_{Phys}(X|m_\nu) ,
\end{eqnarray*}
where
\begin{itemize}
\item $\tilde{X}$ is the reconstructed data point coordinate: $(\tilde{M},\tilde{E})$;
\item $f_{Phys}(X| m_\nu)$ represents the physics, and is a convolution over the
 differential decay rate $d\Gamma(m_\nu,X)/dq^2$, and corrections due to 
 initial state radiation, and beam energy smearing;
\item $G(X,\tilde{X}, \sigma_{data}) $, a detector smearing function,  
 is the probability of an observed $\tilde{X}$ given a true $X$ and
 an estimated error $\sigma_{data}$ ;
\item $\epsilon$ is the total detector efficiency.
\end{itemize}
 Noting that Monte Carlo events are naturally generated with
 a probability density proportional to $f_{Phys}$, 
 the signal likelihood  for a given data event  is rewritten as :
\begin{eqnarray*}
  {\cal L}_{sig}( \tilde{X}_{data}|\sigma_{data} , m_\nu ) = 
{ \sum_{MC | m_\nu=0 } 
	G(\tilde{X}_{data},X_{MC}, \sigma_{data} )
        \times {\cal W}(m_\nu | X_{MC})  \over{{\cal N}(m_\nu )}}.
\end{eqnarray*}
\noindent The sum is taken over all accepted Monte Carlo events, generated
 with a massless neutrino. 
 The term $ {\cal W}(m_\nu | X_{MC}) $ represents the
 relative weight of the Monte Carlo event for a massive neutrino
 compared to a massless neutrino. It is calculated
 at the generated $(M,E)$ rather than the 
reconstructed $(\tilde{M},\tilde{E})$, and  is written as:
\begin{eqnarray*}
{\cal W}(m_\nu | X_{MC}) = 
       { d\Gamma \left (m_\nu |X_{MC}\right) 
	 \over{ d\Gamma \left (m_\nu=0 |X_{MC}\right) }} 
            \Theta \left( X_{MC} \right).
\end{eqnarray*}
\noindent  This is simply the ratio of differential decay rates of a
 tau at the
 (generated) value of $q^2$ appropriate for this Monte Carlo event 
 and this
 value of neutrino mass relative to a massless neutrino. Note that the
 matrix element factorizes and that the hadronic currents cancel in
 this ratio.
 The function $\Theta$ ensures that only Monte Carlo events which
 remain physical at the new neutrino mass are used, and is implemented
 by checking that the scaled energy  of the Monte Carlo event falls
 into the physically allowed interval:
\begin{eqnarray*}
            \Theta \left( X_{MC} \right) = 
                \left\{ \begin{array}{ll}
                    1   &  y_-(m_\nu) \le {E_X\over{E_{B}}} \le y_+(m_\nu) \\
                    0   & else   
                        \end{array}.
                 \right.  
\end{eqnarray*}
 \noindent The kinematic limits are given by:
\begin{eqnarray*}
	y_{\pm} = {1\over{2 M_\tau^2}} (  E_\tau/E_{B} )
                 \left( M_\tau^2+q^2-m_\nu^2 \pm
                          \left( 1 - M_\tau^2/E_\tau^2\right)^{1\over{2}}
		 \lambda^{1\over{2}}\left(M_\tau^2,q^2,\nu^2 \right)
				\right )  , 
\end{eqnarray*}
\vskip -0.5cm
\begin{eqnarray*}
 \lambda(x,y,z) = x^2+y^2+z^2-2(xy+yz+xz). 
\end{eqnarray*}

 The factor ${\cal N}(m_\nu )$ ensures
 that the probability density used in the event likelihood 
 is normalized  for any neutrino mass:
\begin{eqnarray*}
{\cal N}(m_\nu ) 
 & = &
 \sum_{MC  | m_\nu=0 } 
       \left(\int_\Omega G(\tilde{X},X_{MC}, \sigma) d\tilde{X} \right)
 {\cal W}(m_\nu | X_{MC}) .
\end{eqnarray*}

The likelihood function used has been tested on Monte Carlo 
 samples of decays with both 
massless and massive neutrinos and reproduces the input neutrino mass
with no apparent bias.

\end{document}